\documentclass[aps,prmaterials,reprint,amsmath,amssymb,floatfix]{revtex4-2}
\usepackage[T1]{fontenc}
\usepackage{graphicx}
\usepackage{color}
\usepackage{bm}
\usepackage{accents}
\usepackage{siunitx}

\usepackage{tabularx,booktabs,multirow,siunitx}

\usepackage{xr-hyper}
\usepackage[colorlinks,linkcolor=blue,citecolor=blue,urlcolor=blue]{hyperref}
\usepackage{cleveref}

\crefname{equation}{Eq.}{Eqs.}
\Crefname{equation}{Equation}{Equations}
\crefname{figure}{Fig.}{Figs.}
\Crefname{figure}{Figure}{Figures}
\crefname{section}{Sec.}{Secs.}
\Crefname{section}{Section}{Sections}
\crefname{table}{Table}{Tables}
\Crefname{table}{Table}{Tables}
\crefname{appendix}{Appendix}{Appendixes}
\Crefname{appendix}{Appendix}{Appendixes}

\newcommand{\dF}{\Delta {\bm F}}
\newcommand{\tdF}{\Delta \tilde{\bm F}}
\newcommand{\dQ}{\Delta {\bm Q}}
\newcommand{\tdQ}{\Delta \tilde{\bm Q}}
\newcommand{\etajq}{{\bm \eta}_{j {\bm q}}}
\newcommand{\nujq}{{\bm \nu}_{j {\bm q}}}
\newcommand{\tjq}{\tau_{j {\bm q}}}
\newcommand{\qpt}{${\bm q}$-point}
\newcommand{\bvk}{Born-von K{\'a}rm{\'a}n}

\begin{document}

\title{Dilute-Limit Defect Displacements Enabled by Brillouin-Zone Sampling}

\author{Mark E. Turiansky}
\email{mark.e.turiansky.civ@us.navy.mil}
\affiliation{US Naval Research Laboratory, 4555 Overlook Avenue SW, Washington, DC 20375, USA}

\author{John L. Lyons}
\affiliation{US Naval Research Laboratory, 4555 Overlook Avenue SW, Washington, DC 20375, USA}

\author{Noam Bernstein}
\affiliation{US Naval Research Laboratory, 4555 Overlook Avenue SW, Washington, DC 20375, USA}

\date{\today}

\begin{abstract}
    Defects and their interaction with the semiconductor host lattice play an essential role in a variety of technologies.
    When a defect transitions between two electronic states, the lattice distorts in response.
    Large displacements occur near the defect, inducing displacements on neighboring atoms, and so on, producing small displacements on atoms hundreds of {\AA}ngstr{\"o}ms from the defect.
    Describing these displacements accurately is challenging for first-principles calculations due to limited supercell sizes.
    Here we demonstrate a procedure to efficiently obtain dilute-limit atomic displacements, using information obtained in typical modest supercells.
    In our approach, the atomic displacements are first obtained in a small supercell and converted into a force difference under a harmonic potential energy surface.
    The force difference and phonon modes at different {\qpt}s are then unfolded into the {\bvk} supercell to obtain the dilute-limit atomic displacements.
    We critically analyze the convergence behavior of the force difference and study possible electron density differences that give rise to those forces, arguing that modest supercells are sufficient for internal transitions and bound-exciton transitions.
    Two example applications of our approach are given: (1) we calculate the luminescence spectrum of the NV center in diamond and the T center in Si and (2) we obtain dilute-limit configuration coordinate diagrams for these defects.
    In particular, we find that coupling to acoustic phonon modes reduces the accepting-mode frequency in the configuration coordinate diagram.
    Our work provides the missing ingredients necessary to address truly dilute-limit transitions at defects.
\end{abstract}

\maketitle

\section{Introduction}
\label{sec:intro}
Point defects and impurities govern many of the electrical~\cite{pantelides_electronic_1978,mccluskey_dopants_2018} and optical~\cite{stoneham_theory_1975,davies_optical_1989,zhang_ab_2022} properties of semiconductors and insulators.
More recently, defects are being explored~\cite{zhang_material_2020,turiansky_telecom-wavelength_2023,bulancea-lindvall_chlorine_2023,thomas_substitutional_2024} as the basis of quantum technologies including computing, communication, and sensing~\cite{weber_quantum_2010,bassett_quantum_2019-1,wolfowicz_quantum_2021}.
First-principles calculations, particularly those based in density functional theory (DFT), have become an indispensable tool for characterizing and predicting the behavior of defects~\cite{freysoldt_first-principles_2014,dreyer_first-principles_2018,ivady_first_2018,seo_first-principles_2024,turiansky_frontiers_2026}.
Typical calculations are performed in a supercell containing a few hundred atoms within periodic boundary conditions, with the goal of describing an isolated defect in the dilute limit.
Reaching this limit requires overcoming the contamination from interaction with periodic images of the defect, a condition that can be difficult to satisfy and which underlies a long-standing effort to develop so-called finite-size corrections~\cite{lany_assessment_2008,freysoldt_fully_2009,komsa_finite-size_2012,freysoldt_first-principles_2014,gake_finite-size_2020,falletta_finite-size_2020,kumagai_finite-size_2023}.

Many of the relevant physical processes of a defect involve a transition between two electronic states, for example in nonradiative carrier capture~\cite{stoneham_theory_1975,huang_theory_1950,henry_nonradiative_1977,alkauskas_first-principles_2014,shi_comparative_2015,nichols_multiphonon_2025} or luminescence~\cite{stoneham_theory_1975,huang_theory_1950,alkauskas_first-principles_2014-1}.
Corrections to the transition energy to account for interactions with periodic images follow directly from standard finite-size corrections~\cite{freysoldt_fully_2009,komsa_finite-size_2012}.
These finite-size corrections were extended to account for differences in screening that arise in polar materials, giving rise to vertical charge-state corrections~\cite{gake_finite-size_2020,falletta_finite-size_2020}.
Based on this physics, it was later suggested that the curvature of the potential energy surface (PES) for a given transition should also be corrected~\cite{kumagai_finite-size_2023}.

Energetic corrections have thus been thoroughly considered.
However, the atomic displacements between the ground and excited state also converge slowly with supercell size (see, e.g., Ref.~\onlinecite{kumagai_finite-size_2023}), and their dilute-limit scaling has received comparatively little attention.
This slow convergence is rooted in the long-range nature of atomic displacements:
large displacements on atoms near the defect induce displacements on neighboring atoms, and so on.
This effect is most clear in the coupling to acoustic modes observed in luminescence spectra, which is particularly sensitive to small residual displacements far from the defect.
Indeed, coupling to low frequency acoustic modes is absent in typical supercells used in the study of defects~\cite{alkauskas_first-principles_2014-1}.
This issue was addressed with the embedding approach of Alkauskas \textit{et al.}~\cite{alkauskas_first-principles_2014-1}, later refined by Razinkovas \textit{et al.}~\cite{razinkovas_vibrational_2021}.
The embedding approach, while powerful, requires construction and diagonalization of exceptionally large dynamical matrices.

In this work, we show that Brillouin-zone sampling provides a direct and efficient route to dilute-limit defect displacements.
Building upon insight from embedding (\cref{sec:embed}), atomic displacements calculated in modest supercells are converted into a force difference.
Here we critically analyze this force difference within periodic boundary conditions for the three transition types commonly encountered---internal transitions, changes in charge state, and bound-exciton transitions---and connect the force difference's behavior to the underlying charge-density redistribution.
Our analysis provides justification for the short-ranged nature of this force difference, which was assumed in previous work.
Phonon modes are evaluated throughout the Brillouin zone and, along with the forces, are unfolded into the {\bvk} (BvK) supercell.
This procedure converges to the dilute limit far more rapidly than embedding, resolving displacements on atoms hundreds of {\AA}ngstr\"oms from the defect.
We then apply the formalism to compute the luminescence lineshapes of the NV center in diamond and the T center in silicon, and to extract dilute-limit configuration coordinate diagrams (CCDs), which contain the essential ingredients for evaluating nonradiative capture rates.
Our approach is similar to the one used in a recent preprint to study NV center luminescence~\cite{santha_first-principles_2026}, which we became aware of during preparation of this manuscript.

This paper is organized as follows.
In \cref{sec:bkgnd}, we introduce basic definitions related to the atomic displacement vector and phonon modes of the studied system.
In \cref{sec:embed}, we review the embedding approach, which provides useful context for our work.
In \cref{sec:unfold}, we present the main methodology of force unfolding, which incorporates {\qpt} sampling of the Brillouin zone.
We then examine how the force difference behaves within periodic boundary conditions in \cref{sec:forces}, across the three transition types (internal, charge-state, and bound-exciton transitions).
Next, we apply our approach to the case of luminescence in \cref{sec:lum} and dilute-limit CCDs in \cref{sec:ccd}.
Finally, in \cref{sec:discuss} we discuss our results, approximations, and promising future directions, followed by concluding remarks in \cref{sec:concl}.

As our test cases, we will use the NV center in diamond and the T center in Si.
The NV center is the prototype defect being used for applications in quantum information science~\cite{gali_ab_2019,dreyer_first-principles_2018}, and the T center is a burgeoning alternative~\cite{bergeron_silicon-integrated_2020,higginbottom_memory_2023,dhaliah_first-principles_2022}.
We will use the T center for the demonstrations in \crefrange{sec:embed}{sec:forces}.
Details of the DFT calculations are given in \cref{app:comp_det}.

\section{Basic Definitions}
\label{sec:bkgnd}
A change in electronic configuration between a ground and excited state at a defect results in a change in energy $\Delta E$.
As a result of electron-phonon coupling, the change in electronic configuration is also accompanied by a change in the atomic geometry.
This geometry change is quantified through the mass-weighted atomic displacement vector $\dQ$:
\begin{equation}
    \label{eq:dq}
    {\Delta Q}_{I\alpha} = \sqrt{M_I} \left( {R}^0_{e;I\alpha} - {R}^0_{g;I\alpha} \right) \;,
\end{equation}
where $I$ indexes the $N$ atoms and $\alpha$ the three Cartesian directions, $M_I$ is the atomic mass of the $I$th site, and ${\bm R}_{g/e}$ are the atomic coordinates of the ground ($g$) and excited ($e$) states with a superscript zero denoting the equilibrium geometry.
It is the efficient description of $\dQ$ in the dilute limit that is the focus of this work.

\begin{figure}[htb!]
    \centering
    \includegraphics[width=\columnwidth,height=0.3\textheight,keepaspectratio]{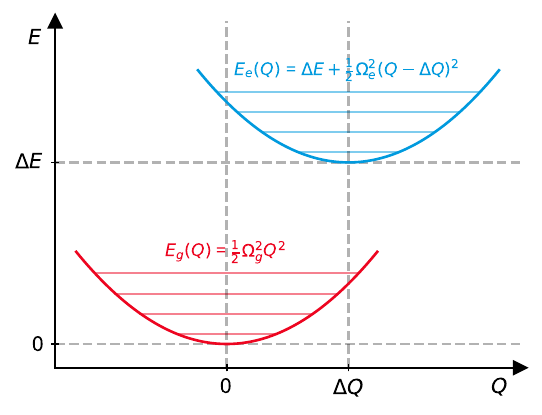}
    \caption{\label{fig:ccd}
        A configuration coordinate diagram showing the energy difference $\Delta E$, magnitude of the atomic displacement $\Delta Q$, and potential energy surface of the ground (red) and excited states (blue), which are characterized by vibrational frequencies $\Omega_{g/e}$.
        Horizontal lines denote the harmonic oscillator levels.
    }
\end{figure}

In addition to $\Delta E$ and $\dQ$, we need a description of the PES that each state experiences.
Under a one-dimensional approximation, the PES can be represented by a CCD (\cref{fig:ccd}).
In the CCD, the depicted vibrational mode connects the equilibrium geometries and is known as the accepting mode~\cite{stoneham_theory_1975,stoneham_non-radiative_1981,stoneham_non-radiative_1978}, which has been used widely as the basis of first-principles calculations of nonradiative decay and optical spectra in cases of strong electron-phonon coupling~\cite{alkauskas_tutorial:_2016,turiansky_frontiers_2026}.
The accepting mode is parallel to $\dQ$ by definition, and therefore, the displacement between the PESs along this direction is simply the norm $\Delta Q = \lVert \dQ \rVert$.
There is an accepting-mode frequency associated with the ground $\Omega_g$ and excited $\Omega_e$ state.
In practice, $\Omega_{g/e}$ is obtained by a finite-difference evaluation along $\dQ$.

The true PES is multidimensional, with three degrees of freedom for each of the $N$ atoms.
In a harmonic approximation, the PES is characterized by the dynamical matrix ${\bm \Phi}$, given by
\begin{equation}
    \label{eq:dynmat}
    \Phi_{I\alpha,J\beta} ({\bm 0}) = \frac{1}{\sqrt{M_I M_J}} \frac{\partial^2 E}{\partial R_{I\alpha} \, \partial R_{J\beta}} \bigg\vert_{{\bm R}^0} \;,
\end{equation}
where $E$ is the total energy of the ground or excited state, and the zero in parentheses indicates that the dynamical matrix corresponds to the $\Gamma$ point of the Brillouin zone.
Generally, ${\bm \Phi}$ differs between the ground and excited state, and ${\bm \Phi}$, $E$, and ${\bm R}$ should have a ``$g$/$e$'' label.
An equal-mode approximation (${\bm \Phi}_g \! = \! {\bm \Phi}_e$) has been quite effective for luminescence calculations~\cite{alkauskas_first-principles_2014-1}.
However, the differing ${\bm \Phi}$'s also manifests as a difference in $\Omega_g$ and $\Omega_e$, which is important in the context of nonradiative decay~\cite{alkauskas_first-principles_2014,turiansky_nonrad_2021,stoneham_theory_1975}.
We have assumed the equal-mode approximation here for notational simplicity, dropping the ``$g/e$'' labels except in cases of ambiguity.
All quantities will be taken from the ground state, unless otherwise stated, and a discussion on relaxing this approximation is found in \cref{sec:discuss}.

While ${\bm \Phi} ({\bm 0})$ encompasses the 3$N$ degrees of freedom within the cell, it does not fully account for periodic boundary conditions.
When the effective range of interactions is captured within the supercell, the dynamical matrix at an arbitrary point ${\bm q}$ can be obtained by Fourier interpolation:
\begin{equation}
    \label{eq:dynmat_q}
    \Phi_{I\alpha,J\beta} ({\bm q}) = e^{i {\bm q} \cdot ({\bm R}^0_I - {\bm R}^0_J)} \, \Phi_{I\alpha,J\beta} ({\bm 0}) \;.
\end{equation}
The difference in atomic positions that enters the phase factor should be understood as the shortest vector within periodic boundary conditions.
When multiple equivalent shortest vectors exist, we average over their phases, following previous works~\cite{togo_implementation_2023,parlinski_first-principles_1997}.

The phonon eigenvectors and frequencies are then obtained by diagonalizing ${\bm \Phi} ({\bm q})$,
\begin{equation}
    \label{eq:diag}
    {\bm \Phi}({\bm q}) \, {\bm U}({\bm q}) = {\bm U}({\bm q}) \, {\bm \Omega}^2({\bm q}) \;,
\end{equation}
where ${\bm U}({\bm q}) = ({\bm \eta}_{1 {\bm q}} \, \dots \, {\bm \eta}_{3N {\bm q}})$ is the unitary transformation matrix with the eigenvectors $\etajq$ as columns.
${\bm \Omega}^2({\bm q}) = {\rm diag}(\omega_{1 {\bm q}}^2, \, \dots, \, \omega_{3N {\bm q}}^2)$ is the diagonal matrix of eigenvalues, which correspond to the phonon frequencies $\omega_{j {\bm q}}$ squared.

We can examine the PES from the view of normal modes by rotating the atomic displacement vector $\dQ$ into the phonon basis.
In this view, each $\Gamma$-point phonon mode contributes an amount $\tjq$ to the atomic displacement:
\begin{equation}
    \label{eq:tau_0}
    \tau_{j {\bm 0}} = {\bm \eta}_{j {\bm 0}} \cdot \dQ \;.
\end{equation}
We can thus imagine $3N$ CCDs, one for each mode, with energy difference $\Delta E$, frequency $\omega_{j {\bm 0}}$, and atomic displacement $\tau_{j {\bm 0}}$.
This expression is simply a transformation of the atomic displacement vector:
the determination of $\tjq$ in the dilute limit is discussed in the two sections that follow.

\section{Embedding}
\label{sec:embed}
We begin by reviewing the embedding approach that has been previously used to examine dilute-limit defect relaxations for luminescence prediction~\cite{alkauskas_first-principles_2014-1,razinkovas_vibrational_2021,jin_photoluminescence_2021,bouquiaux_first-principles_2023,silkinis_optical_2024,silkinis_optical_2025,bouquiaux_micro-environment_2026}.
The embedding approach not only provides important context for our work, but also motivates an important shift in perspective from displacements to forces, as elaborated below.
Specifically, the atomic displacements are related to forces within a harmonic approximation using the dynamical matrix at the $\Gamma$ point:
\begin{equation}
    \label{eq:dF}
    \dF = {\bm \Phi} ({\bm 0}) \, \dQ \;,
\end{equation}
where $\dF$ is the difference in forces between the ground and excited state evaluated in the same geometry, having units of eV/amu$^{1/2}$~{\AA}.
There is a connection between \cref{eq:dF} and the concept of Kanzaki forces~\cite{kanzaki_point_1957}.
Kanzaki forces describe the inclusion of a defect with respect to the pristine lattice, whereas here we consider an excited state with respect to the ground state.

The shift to a focus on forces is conceptually important, as this force difference is more short-ranged than the displacements.
This asymmetry emerges from the asymmetry between ${\bm \Phi}$ and its inverse ${\bm \Phi}^{-1}$.
Consider the simple case of a material with nearest-neighbor interactions.
Evaluating \cref{eq:dF} for a $\dQ$ on just a few atoms will produce a $\dF$ that is nonzero on only those atoms and their neighbors.
If we instead ask what displacements are produced when $\dF$ is on just a few atoms, we need information on ${\bm \Phi}^{-1}$.
Nearest-neighbor interactions produces an elastic dynamical matrix with ${\bm \Phi} \! \sim \! \lVert {\bm q} \rVert$.
The inverse ${\bm \Phi}^{-1} \! \sim \! {\lVert {\bm q} \rVert}^{-1}$, or by Fourier transforming to real space, ${\bm \Phi}^{-1} \! \sim \! {\lVert r \rVert}^{-2}$.
Thus when $\dF$ is localized on just a few atoms, inverting \cref{eq:dF} produces long-range displacements that decay as ${\lVert r \rVert}^{-2}$~\cite{eshelby_determination_1957,stoneham_theory_1975,lopez-morales_density_2026}.
These arguments do not tell us whether $\dF$ is actually short-ranged in reality---just that it is more short-ranged than $\dQ$;
the localization of $\dF$ can be assessed by examining the underlying electronic structure of the defect and is discussed in detail in \cref{sec:forces}.

Assuming $\dF$ is appropriately short-ranged, embedding proceeds by appending zeros to $\dF$ out to some large target supercell.
We will label quantities in this large supercell with a tilde: e.g., $\tdF$ for the zero-padded force difference.
The tilde also implies $\Gamma$-point sampling, and the ${\bm q}$ index will be suppressed on $\tilde{\bm U}$ and $\tilde{\bm \Omega}$ for the rest of this section.
\Cref{eq:dF} can then be inverted,
\begin{equation}
    \label{eq:tilde_dQ}
    \tdQ = \left[ \tilde{\bm U} \, \tilde{\bm \Omega}^{-2} \, \tilde{\bm U}^\mathsf{T} \right] \, \tdF \;.
\end{equation}
which gives the atomic displacement vector in the large supercell $\tdQ$.

Typically $\tdQ$ does not need to be explicitly evaluated.
Instead we can obtain the per-mode atomic displacement $\tilde{\tau}_j$ by writing \cref{eq:tau_0} in matrix form in the large supercell:
\begin{align}
    \label{eq:tau_emb_mat_0}
    \tilde{\bm \tau} &= \tilde{\bm U}^\mathsf{T} \, \tdQ \;, \\
    \label{eq:tau_emb_mat_1}
    &= \tilde{\bm \Omega}^{-2} \, \tilde{\bm U}^\mathsf{T} \, \tdF \;,
\end{align}
or explicitly with all indices,
\begin{equation}
    \label{eq:tau_emb}
    \tilde{\tau}_{j} = \frac{1}{\tilde{\omega}_{j}^{2}} \sum_{I,\alpha} \tilde{\eta}_{j; I \alpha} \, \Delta \tilde{F}_{I \alpha} \;,
\end{equation}
where we've used the orthogonality of the eigenstates ($\tilde{\bm U}^\mathsf{T} \, \tilde{\bm U} = {\bm 1}$) to reach \cref{eq:tau_emb_mat_1}.
$\tilde{\bm \tau}$ completely determines $\tdQ$ and, conveniently, is also the quantity needed to evaluate luminescence (see \cref{sec:lum}).

To evaluate \cref{eq:tau_emb} and to achieve the inversion in \cref{eq:tilde_dQ}, we need to construct and diagonalize the large supercell dynamical matrix $\tilde{\Phi}$.
This is a sticking point for the embedding methodology.
Converging the atomic displacements often requires supercells containing more than 10,000 atoms~\cite{alkauskas_first-principles_2014-1,razinkovas_vibrational_2021}.
Evaluating the dynamical matrix for a system with that many degrees of freedom is challenging and requires approximations.
Some works have constructed the dynamical matrix by embedding the defect dynamical matrix ${\bm \Phi} ({\bm 0})$ into the dynamical matrix from a pristine supercell calculation~\cite{alkauskas_first-principles_2014-1,razinkovas_vibrational_2021,jin_photoluminescence_2021}, potentially unfolding the {\qpt}-sampled dynamical matrix of the pristine supercell~\cite{bouquiaux_first-principles_2023}.
In previous work, we've demonstrated that MLIPs can be used to obtain the large supercell dynamical matrix, avoiding this \textit{ad hoc} dynamical matrix construction while maintaining accuracy~\cite{turiansky_machine_2026}.
Additionally, one might assume short-ranged interactions to obtain a sparse dynamical matrix, facilitating diagonalization~\cite{razinkovas_vibrational_2021}.

\begin{figure}[htb!]
    \centering
    \includegraphics[width=\columnwidth,height=0.3\textheight,keepaspectratio]{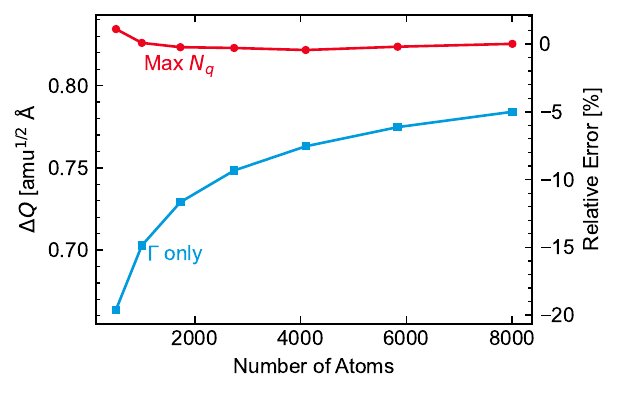}
    \caption{\label{fig:dq_conv}
        The atomic displacement magnitude $\Delta Q$ as a function of the number of atoms in the embedded supercell size.
        Results for $\Gamma$ point-only sampling (discussed in \cref{sec:embed}) are shown in blue and for the largest {\qpt} grid (discussed in \cref{sec:unfold}) in red.
        Error is measured relative to the largest supercell with the max {\qpt} grid.
    }
\end{figure}

\Cref{fig:dq_conv} shows the magnitude of the atomic displacement vector $\Delta Q$ as a function of the embedded supercell size for the T center in Si as obtained for the conventional embedding approach (labeled as ``$\Gamma$ only'').
There is a significant size dependence, slowly converging toward the dilute limit.
We confirm that there are displacements outside the small cell boundaries, reinforcing the importance of embedding forces.
While practical and foundational, the embedding approach exhibits rather slow convergence.

\section{Force Unfolding}
\label{sec:unfold}
We now incorporate {\qpt} sampling beyond the $\Gamma$ point in the definition of $\tjq$.
This is best done by considering the BvK supercell~\cite{born_uber_1912}.
Sampling the Brillouin zone with a $N_q \times N_q \times N_q$ {\qpt} grid is equivalent to $\Gamma$-point sampling of a $N_q \times N_q \times N_q$ multiple of the supercell.
To be clear, the BvK supercell utilized here is a multiple of the \textit{defective supercell}, which is already a multiple of the unit cell.

\begin{figure}[htb!]
    \centering
    \includegraphics[width=\columnwidth,height=0.3\textheight,keepaspectratio]{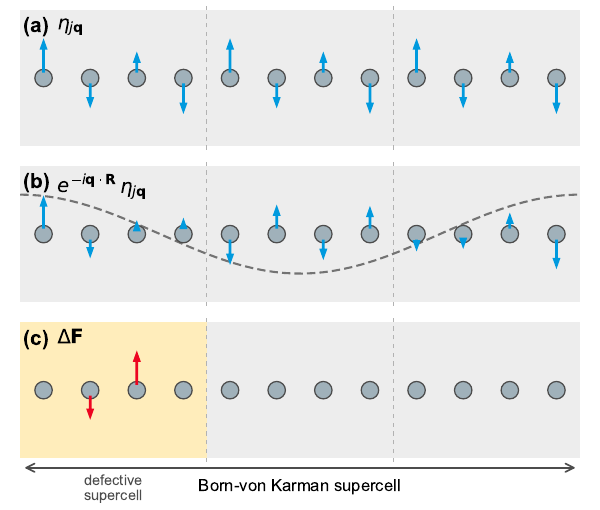}
    \caption{\label{fig:bvk}
        A schematic depiction of unfolding into the {\bvk} supercell.
        (a) The phonon eigenvector $\etajq$ is repeated in each periodic image.
        (b) The ${\bm q}$-dependent phase factor is applied.
        (c) The force difference $\dF$ is set to zero outside one image.
        Atoms are shown as gray circles.
    }
\end{figure}

\Cref{fig:bvk} shows the unfolding of a given eigenvector $\etajq$ into the BvK supercell for a schematic one-dimensional system.
The $3N$-dimensional eigenvector is repeated in each periodic image of the supercell [\cref{fig:bvk}(a)].
Next the ${\bm q}$-dependent phase factor is applied [\cref{fig:bvk}(b)].
In contrast, the force difference is assumed to be non-zero in only one image of the BvK supercell and zero elsewhere [\cref{fig:bvk}(c)].
No phase factor is needed.
Because the force difference is non-zero in only one image, all quantities can be obtained within the defective supercell without explicitly constructing the BvK supercell.

It is useful to define the auxiliary phase-including eigenvector $\nujq$, given by
\begin{equation}
    \label{eq:nu_q}
    \nu_{j {\bm q}; I \alpha}  = e^{-i {\bm q} \cdot {\bm R}^0_{g;I}} \, \eta_{j {\bm q}; I \alpha} \;.
\end{equation}
We note that there are several instances of arbitrariness in the phase definition:
(1) there is the inherent arbitrary phase of $\etajq$ from numerical diagonalization;
(2) there is the definition of the origin for the atomic coordinates ${\bm R}_{g;I}^0$;
and (3) there is the choice of the image in the BvK supercell where the force difference is non-zero.
In all observables where $\nujq$ is used, the quantity of interest either uses the magnitude or time-reversal symmetry mitigates the phase uncertainty.
As a result, the arbitrariness is not a concern.

Based on the BvK unfolding, we propose the following expression for $\tjq$,
\begin{equation}
    \label{eq:tau_q}
    \tjq = \frac{1}{\omega_{j {\bm q}}^{2}} \sum_{I,\alpha} \nu_{j {\bm q}; I \alpha}^* \, \Delta F_{I \alpha} \;,
\end{equation}
which is obtained by replacing $\etajq$ with $\nujq$ in \cref{eq:tau_emb}.
Alternative forms of $\tjq$ are considered in \cref{app:alt} with an explanation of why they are non-optimal.
\Cref{eq:tau_q} can be put in matrix form by introducing the unitary matrix,
\begin{equation}
    \label{eq:V_q}
    {\bm V} ({\bm q}) = ({\bm \nu}_{1 {\bm q}} \dots {\bm \nu}_{3N {\bm q}}) \;,
\end{equation}
which is the phase-including counterpart of ${\bm U} ({\bm q})$.
We thus obtain
\begin{equation}
    \label{eq:tau_q_mat}
    {\bm \tau} ({\bm q}) = {\bm \Omega}^{-2} ({\bm q}) \, {\bm V}^\dagger ({\bm q}) \, \dF \;,
\end{equation}
as the matrix form.

\begin{figure}[htb!]
    \centering
    \includegraphics[width=\columnwidth,height=0.3\textheight,keepaspectratio]{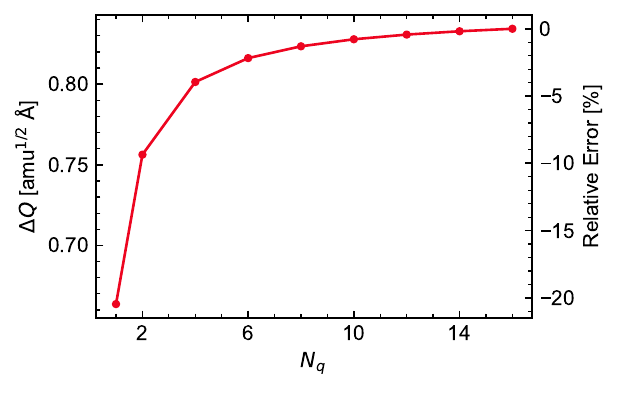}
    \caption{\label{fig:qpt_conv}
        The atomic displacement magnitude $\Delta Q$ as a function of the number of grid points $N_q$ in the $N_q \times N_q \times N_q$ {\qpt} grid in the 512-atom supercell.
        Error is measured relative to the largest $N_q$.
    }
\end{figure}

From the mode-resolved atomic displacements $\tjq$, the magnitude of the atomic displacement $\Delta Q$ is given by
\begin{equation}
    \label{eq:dilute_dq}
    \Delta Q^2 = \sum_{j,{\bm q}} w_{\bm q} \, {\lvert \tjq \rvert}^2 \;,
\end{equation}
where $w_{\bm q}$ is the weight of the given {\qpt} after considering symmetry.
\Cref{fig:qpt_conv} shows the convergence of $\Delta Q$ as a function of the {\qpt} grid.
We observe rapid convergence to the dilute limit with increasing grid size.

In addition to converging $\dQ$, embedding has the potential advantage of providing a better description of the dilute-limit phonons because the embedded cell is not perturbed by images of the defect like the BvK supercell.
(This comes at the expense of a more costly dynamical-matrix diagonalization.)
To assess the extent to which the phonon description impacts the results, we test the convergence of our procedure in combination with embedding.
We note that both embedding and our procedure relies on padding $\dF$ with zeros: in embedding, this is explicit, while in our procedure it is implicit.
In this test, we first embed the force difference into a larger supercell, following \cref{sec:embed}, and then we apply \cref{eq:tau_q} on a converged {\qpt} grid.
Because the force difference is identical in every calculation, the only parameter that changes is the description of the phonon modes, which isolates its effect on the convergence of $\Delta Q$.
The results are shown by the red line in \cref{fig:dq_conv}.
Even for the smallest supercell containing 512 atoms, $\Delta Q$ is within $\approx$1\% of the large-cell (8000 atoms) value when using {\qpt} sampling with $N_q = 16$.
For comparison, using $\Gamma$-only sampling for the small supercell produced a value $\approx$20\% smaller than the same {\qpt} sampled, large-cell value.
We conclude that the effect of converging the phonon modes is relatively minor.

While not explicitly necessary for many calculations, it can be useful to examine the full displacement pattern in the BvK supercell.
We introduce an index ${\bm \ell} = (l_1, l_2, l_3)$ for each image of the BvK supercell.
The BvK $\dQ$ is then obtained through a Fourier transform of the per-mode displacements after transforming back to real space:
\begin{equation}
    \label{eq:big_dq}
    \Delta Q^{\rm BvK}_{{\bm \ell} I \alpha} = \sum_{j, {\bm q}} e^{i {\bm q} \cdot {\bm r}_{\bm \ell}} \, \nu_{j {\bm q}; I \alpha} \, \tjq \;,
\end{equation}
where ${\bm r}_{\bm \ell} = l_1 {\bm a}_1 + l_2 {\bm a}_2 + l_3 {\bm a}_3$, or in matrix form,
\begin{equation}
    \label{eq:big_dq_mat}
    \dQ^{\rm BvK}_{\bm \ell} = \sum_{\bm q} e^{i {\bm q} \cdot {\bm r}_{\bm \ell}} \, {\bm V} ({\bm q}) \, {\bm \tau} ({\bm q}) \;.
\end{equation}
We note that, as previously mentioned, the phase of $\nujq$ compensates the phase of $\nujq^*$ within $\tjq$ [\cref{eq:tau_q}], and $\dQ^{\rm BvK}_{\bm \ell}$ is real.

Evaluating \cref{eq:big_dq,eq:big_dq_mat} on a symmetry-reduced grid requires care to account for the various phases when unfolding.
For simplicity, we use the full grid without considering point group reductions;
however, we have considered time-reversal symmetry, which is relevant for most systems of interest.
Time-reversal symmetry amounts to taking the real part of \Cref{eq:big_dq}:
\begin{equation}
    \label{eq:big_dq_trs}
    \Delta Q^{\rm BvK}_{{\bm \ell} I \alpha} = \mathrm{Re} \left( \sum_{j, {\bm q}} w_{\bm q} \, e^{i {\bm q} \cdot {\bm r}_{\bm \ell}} \, \nu_{j {\bm q}; I \alpha} \, \tjq \right) \;,
\end{equation}
where the weights $w_{\bm q}$ account for time-reversal symmetry reductions.
Other quantities considered in this work [e.g., \cref{eq:dilute_dq}] only rely on the magnitude, for which the full point group symmetries can be easily applied.

\begin{figure*}[htb!]
    \centering
    \includegraphics[width=\textwidth,height=0.5\textheight,keepaspectratio]{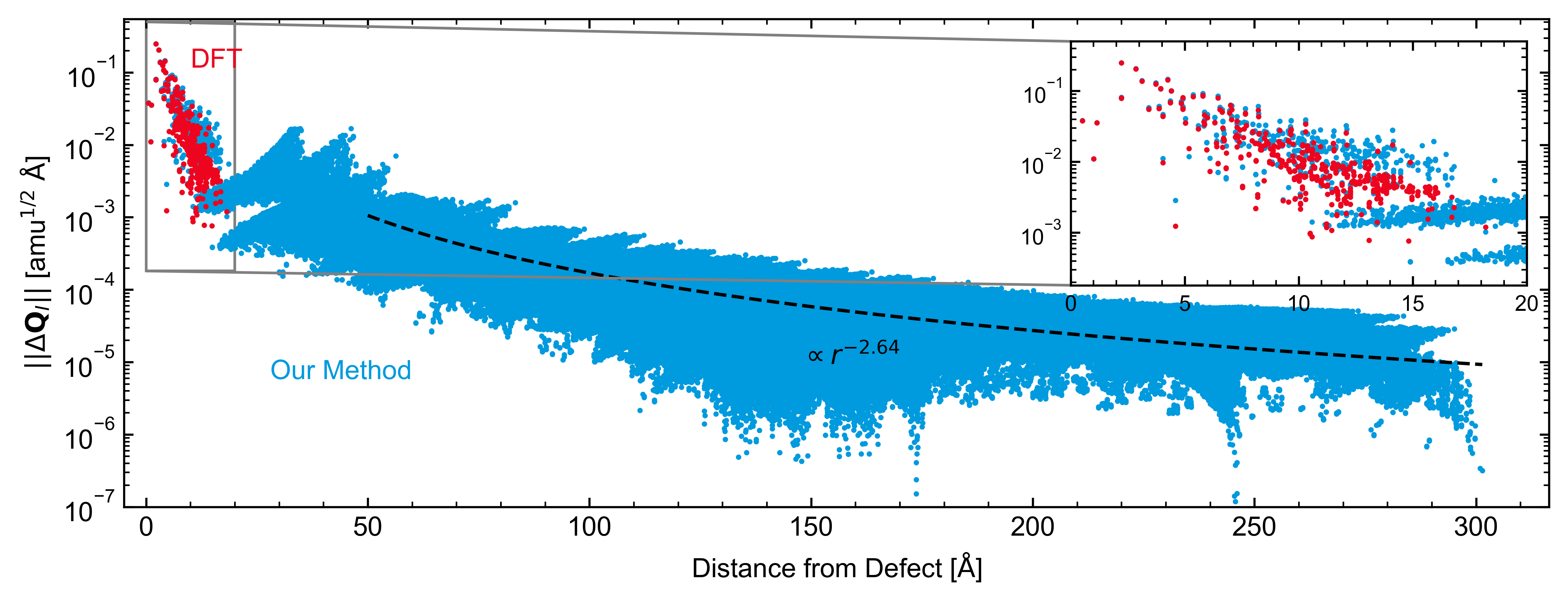}
    \caption{\label{fig:big_dq}
        The displacement magnitude $\lVert \dQ_I \rVert$ on the $I$th atom as a function of distance from the defect from explicit DFT calculations (red) and unfolding into the {\bvk} supercell [blue; \cref{eq:big_dq}].
        The black, dashed line shows a fit of the data in the range of 100 to 174~{\AA} to a power-law decay from which we find an exponent of $\approx$2.64.
        We take the average position of the C and H atoms in the T center as the center of the defect.
    }
\end{figure*}

$\dQ^{\rm BvK}$ compared to $\dQ$ from DFT as a function of distance from the defect is shown in \cref{fig:big_dq}.
For this $16 \times 16 \times 16$ {\qpt} grid, there are more than 2 million atoms in the BvK supercell. 
Immediately we can see that our procedure allows us to probe displacements on atoms 300~{\AA} from the defect.
This is compared to probing displacements $<$20~{\AA} from the defect in the DFT calculation or $<$100~{\AA} with embedding~\cite{razinkovas_vibrational_2021}.
From the inset of \cref{fig:big_dq}, we see that both $\dQ$ values agree for atoms $<$5~{\AA} from the defect.
On the other hand, we find that periodic boundary conditions impose an artificial constraint on the displacement of atoms near the boundary of the defect supercell, which displace further in the BvK supercell.
Three peaks in the displacement pattern also appear on atoms 30--50~{\AA} from the defect.
The long-range atomic displacements exhibit a power-law decay (black, dashed line in \cref{fig:big_dq}) with an exponent of $\approx$2.64, suggesting a strain field beyond a simple volumetric inclusion~\cite{eshelby_determination_1957,stoneham_theory_1975}.
This exponent was obtained by fitting data in the range of 100 to 174~{\AA} (half the size of the BvK supercell), to exclude near-field effects and the region contaminated by periodic images.

We have demonstrated that our approach is a practical way to drastically accelerate the evaluation of defect displacements in the dilute limit.
It is also efficient: parallelization over independent {\qpt}s is straightforward, and the diagonalized dynamical matrices are small, requiring only modest memory and computational resources.
Moreover, our approach resolves atomic displacements on the order of $10^{-6}$~amu$^{1/2}$~{\AA}.
Evaluating such small displacements explicitly with DFT is practically impossible, requiring exceptionally small energy and force convergence criteria.
To complete the justification of our approach, we address in the next section whether the force difference $\dF$ is convergent in small supercells that can be addressed with DFT.
We note that this justification not only applies to our formulas here but also to the embedding approach, which originally motivated the use of forces.

\section{Forces Within Periodic Boundary Conditions}
\label{sec:forces}
$\dF$ should be interpreted as a force difference between the ground and excited state at the same geometry; i.e.,
\begin{equation}
    \label{eq:dF_def}
    \Delta F_{I\alpha} = F_{e;I\alpha} ({\bm R}_0) - F_{g;I\alpha} ({\bm R}_0) \; ,
\end{equation}
where
\begin{equation}
    \label{eq:F_def}
    F_{g/e;I\alpha} ({\bm R}_0) = -\frac{1}{\sqrt{M_I}} \frac{\partial E_{g/e}}{\partial R_{g/e;I\alpha}} \bigg\vert_{{\bm R}_0}
\end{equation}
and ${\bm R}_0$ is the atomic coordinates at which the forces ${\bm F}_{g/e}$ are evaluated.
For a harmonic PES, the forces point to their respective equilibrium geometry, giving the correct $\dQ$ [\cref{eq:dq}] independent of the geometry at which the forces are evaluated.

Within density functional theory, the atomic forces are obtained from the Hellmann-Feynman theorem~\cite{feynman_forces_1939}, requiring derivatives of the external (electron-ion) and ion-ion potentials with respect to atomic displacements.
Since both terms of $\dF$ are evaluated in the same geometry [\cref{eq:dF_def}], the ion-ion terms cancel.
For an external potential produced by the ions in the absence of additional perturbations, we have
\begin{equation}
    \label{eq:dF_dft}
    \Delta F_{I\alpha} = -\frac{Z_I}{\sqrt{M_I}} \frac{\partial}{\partial R_{I\alpha}} \int_V d{\bm r} \, \frac{\delta n({\bm r})}{\lvert {\bm r} - {\bm R}_I \rvert} \;,
\end{equation}
where $Z_I$ is the charge of ion $I$, $V$ is the supercell volume, and $\delta n({\bm r}) = n_e({\bm r}) - n_g({\bm r})$ is the difference in electron density between the ground and excited states.
These forces arise from the interaction of the dipole produced by displacing the point charge $Z_I$ of ion $I$ with the fields produced by $\delta n$.

Therefore, the convergence properties of $\dF$ can be understood by considering $\delta n$ in detail.
$\delta n$ is derived from the orbitals that change occupation during the transition.
Deep-defect states are localized on the scale of a bond length.
Shallow-defect or bound-exciton states involve hydrogenic wavefunctions, which are similarly localized---albeit on a length scale that does not fit in typical supercells, requiring care in actual calculations.
In the dilute limit, the $\delta n$ arising from a transition at a defect should therefore be exponentially localized on the defect.

Assuming $\delta n ({\bm r})$ is sufficiently localized, we propose that \cref{eq:dF_dft} can be re-evaluated in open boundary conditions (taking $V$ to infinity) to obtain the dilute-limit force difference.
Direct implementation of this idea with real electron densities is complicated by the use of pseudopotentials, non-local interactions, \textit{etc}.
Instead, we now use model densities to build understanding for three distinct cases of $\delta n({\bm r})$ that are commonly encountered.

\textit{Case I: Internal Transitions--- }
In an internal transition, an electron transitions between two deep defect orbitals.
The defect's charge state is unchanged.
We therefore have the requirement that
\begin{equation}
    \label{eq:dn_eq_0}
    \int_V d{\bm r} \, \delta n({\bm r}) = 0 \;.
\end{equation}
For internal transitions, $\delta n$ has no monopole moment, producing at most a dipolar electric field.

To study this case, we construct a model density from Gaussian point charges,
\begin{equation}
    \label{eq:gaussian}
    g({\bm r}; {\bm \mu}, \sigma) = \frac{1}{\mathcal{N}} e^{-\lvert{\bm r} - {\bm \mu}\rvert^2 / 2 \sigma^2} \;,
\end{equation}
where the normalization constant is given by
\begin{equation}
    \label{eq:norm_const}
    \mathcal{N}^{-1} = \int_V d{\bm r} \, e^{-\lvert{\bm r} - {\bm \mu}\rvert^2 / 2 \sigma^2} \;.
\end{equation}
We define the model density
\begin{equation}
    \label{eq:dip_model}
    \delta n_m ({\bm r}) = g({\bm r}; -{\bm r}_d / 2, \sigma_d) - g({\bm r}; {\bm r}_d / 2, \sigma_d) \;,
\end{equation}
where ${\bm r}_d = (0, 0, 1)$~{\AA} and $\sigma_d = 0.5$~{\AA}.
In practice, $\delta n_m$ is represented on a $200 \times 200 \times 200$ real-space grid.

\begin{figure}[htb!]
    \centering
    \includegraphics[width=\columnwidth,height=0.3\textheight,keepaspectratio]{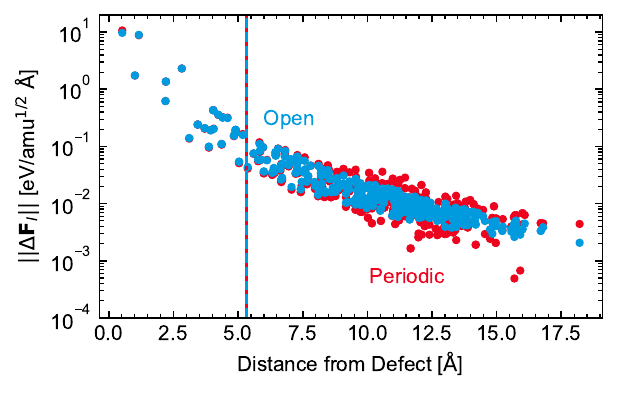}
    \caption{\label{fig:obc_I}
        Model density corresponding to an internal transition: the force difference magnitude $\lVert \dF_I \rVert$ on atom $I$ as a function of distance from the center of charge.
        Periodic boundary conditions are shown in red and open boundary conditions in blue.
        Vertical lines denote the radius that contains 99\% of the force difference magnitude.
    }
\end{figure}

We evaluate \cref{eq:dF_dft} with $\delta n_m$ in both periodic and open boundary conditions, as shown in \cref{fig:obc_I}.
For the ions in \cref{eq:dF_dft}, we use those of the T center in Si, with the origin set to the average position of the C and H atoms in the T center.
We find that atoms near the defect have the same force difference independent of boundary conditions.
Further from the defect, some differences arise, but the overall envelope of the force decay is preserved.
For both boundary conditions, 99\% of the force difference magnitude $\lVert \dF \rVert$ is contained within 5.3~{\AA} of the defect, as shown by the vertical lines in \cref{fig:obc_I}.

Based on these findings, we suggest that the force difference associated with internal transitions is likely sufficiently converged within typical supercells and can be used to extrapolate the displacements to the dilute limit.
We note that the dipolar model density is a ``worst-case scenario'' for the localization range of the force difference associated with an internal transition.
Some transitions may have a negligible dipole moment and be dominated by quadrupole or higher terms, which are less long range.
This is all consistent with the empirical observation that the Alkauskas formulation of luminescence based on embedding has been successfully applied to various internal transitions of defects~\cite{alkauskas_first-principles_2014-1,razinkovas_vibrational_2021,jin_photoluminescence_2021,turiansky_machine_2026,bouquiaux_first-principles_2023}.

\textit{Case II: Changes in Charge State--- }
During a capture or emission process, an electron is exchanged between a deep defect state and the delocalized bulk states of the host material.
The defect changes its charge state.
Calculations of a defect in different charge states involve changing the total number of electrons in the system (with an implicit compensating background charge).
The charge density of the bulk state thus does not enter the evaluation of the forces, and we can model $\delta n$ as a Gaussian point charge:
\begin{equation}
    \label{eq:mono_model}
    \delta n_m ({\bm r}) = g({\bm r}; {\bm 0}, \sigma_d) \;,
\end{equation}
where the same value of $\sigma_d$ is used.
\Cref{eq:dn_eq_0} no longer holds, and instead we have
\begin{equation}
    \label{eq:dn_eq_1}
    \int_V d{\bm r} \, \delta n({\bm r}) = 1
\end{equation}
by definition.

\begin{figure}[htb!]
    \centering
    \includegraphics[width=\columnwidth,height=0.3\textheight,keepaspectratio]{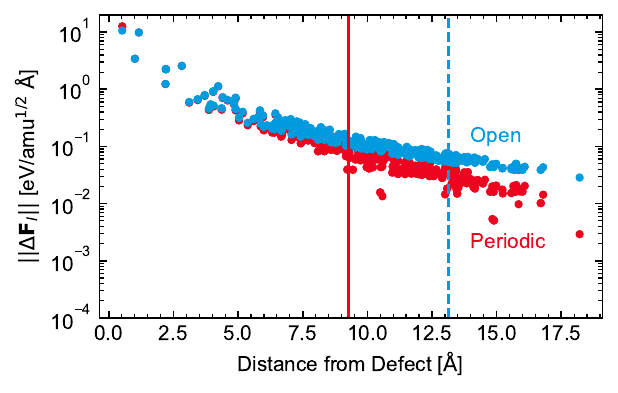}
    \caption{\label{fig:obc_II}
        Model density corresponding to a change in charge state: the force difference magnitude $\lVert \dF_I \rVert$ on atom $I$ as a function of distance from the center of charge.
        Periodic boundary conditions are shown in red and open boundary conditions in blue.
        Vertical lines denote the radius that contains 99\% of the force difference magnitude.
    }
\end{figure}

We again evaluate \cref{eq:dF_dft} with the new $\delta n_m$, as shown in \cref{fig:obc_II}.
While the force difference in periodic and open boundary conditions generally agree for atoms close to the defect, they begin to diverge further away with the periodic force difference being consistently lower.
As observed in the displacement pattern (\cref{fig:big_dq}), there is an artificial suppression on atoms near the boundary due to periodic boundary conditions.
This result suggests that the $\dF$ arising from a change in charge state may converge slower than that of an internal transition:
a field arising from a monopole is longer range than that of a dipole.
We find that 99\% of the force difference magnitude is contained within 9.3~{\AA} of the defect in periodic boundary conditions, but that value increases to 13.1~{\AA} in open boundary conditions.

For transitions involving a change in charge state, we recommend using the largest supercell possible to try to mitigate the force contamination within periodic boundary conditions.
However, this consideration should be taken in context.
Carrier capture or emission processes are most readily observed for defects with strong electron-phonon coupling, which has the effect of washing out details (e.g., producing a featureless phonon sideband resembling a Gaussian in luminescence~\cite{stoneham_theory_1975,alkauskas_tutorial:_2016}).
Subtle changes from force convergence may end up being immaterial for relevant cases, but care should be exercised.
The ideal approach to avoid large supercell calculations would be to re-evaluate the force difference within open boundary conditions for the true $\delta n$, with pseudopotentials and all complications addressed.
This should be addressed in future work.

\textit{Case III: Bound Excitons--- }
A bound-exciton transition involves an electron transiting between a deep defect state and a loosely bound hydrogenic state produced by the Coulomb potential of the nominally charged defect.
Such a transition shares similarities with both of the previously discussed cases.
As in an internal transition, the total charge is conserved, i.e., \cref{eq:dn_eq_0} is valid.
However, as the hydrogenic state is loosely bound and often does not fit in the supercells used to study defects, the defect appears as if it has changed its charge state.
Indeed, using the atomic geometry of the charged defect in place of an explicit calculation of the bound-exciton state has proven effective for calculating luminescence~\cite{silkinis_optical_2025,turiansky_machine_2026}.

To model a bound-exciton transition, we use a model density constructed from two concentric Gaussian point charges:
\begin{equation}
    \label{eq:be_model}
    \delta n_m ({\bm r}) = g({\bm r}; {\bm 0}, \sigma_d) - g({\bm r}; {\bm 0}, \sigma_{BE}) \;,
\end{equation}
where $\sigma_d$ is the same and $\sigma_{BE} = 3.0$~{\AA} is chosen to fit the hydrogenic state within the supercell, allowing us to understand the dilute-limit behavior.
The constructed $\delta n_m$ obeys \cref{eq:dn_eq_0} as required.

\begin{figure}[htb!]
    \centering
    \includegraphics[width=\columnwidth,height=0.3\textheight,keepaspectratio]{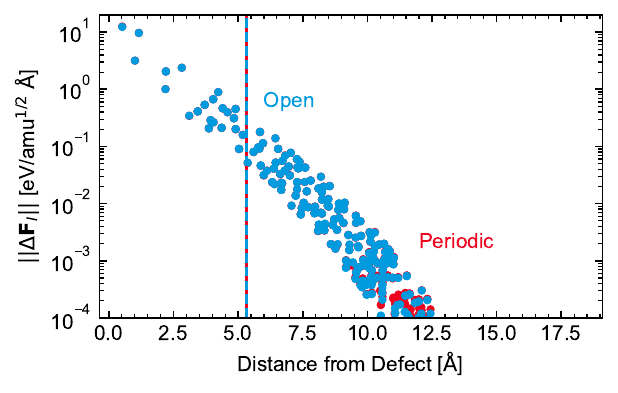}
    \caption{\label{fig:obc_III}
        Model density corresponding to a bound-exciton transition: the force difference magnitude $\lVert \dF_I \rVert$ on atom $I$ as a function of distance from the center of charge.
        Periodic boundary conditions are shown in red and open boundary conditions in blue.
        Vertical lines denote the radius that contains 99\% of the force difference magnitude.
    }
\end{figure}

The force difference from evaluating \cref{eq:dF_dft} with this model density is depicted in \cref{fig:obc_III}.
Here the force difference in periodic and open boundary conditions agree well over a large range with the forces decaying exponentially.
The constructed $\delta n_m$ has no monopole moment and is radially symmetric such that higher-order multipole moments are also zero.
We find that 99\% of the force difference magnitude is (coincidentally) within 5.3~{\AA} of the defect for both boundary conditions.

\begin{figure}[htb!]
    \centering
    \includegraphics[width=\columnwidth,height=0.3\textheight,keepaspectratio]{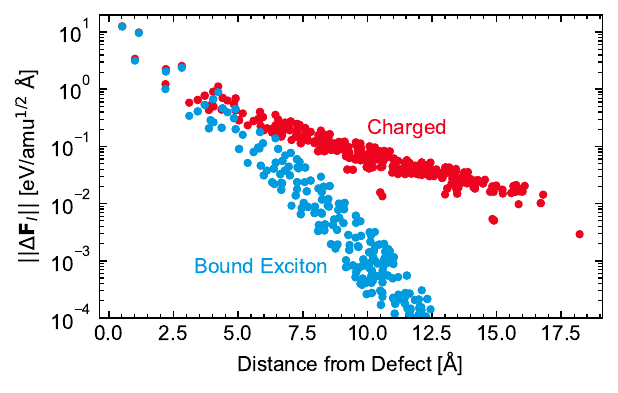}
    \caption{\label{fig:chg_vs_be}
        A comparison of the model densities corresponding to a change in charge state (red) and a bound-exciton transition (blue): the force difference magnitude $\lVert \dF_I \rVert$ on atom $I$ as a function of distance from the center of charge in periodic boundary conditions.
    }
\end{figure}

The exponential decay of the force difference for a bound-exciton transition is encouraging.
As most bound-exciton wavefunctions do not fit in typical supercell sizes, a charged supercell may be used to approximate the bound-exciton state.
In \cref{fig:chg_vs_be}, we re-plot the periodic boundary condition force difference from a change in charge state (Case II; \cref{fig:obc_II}) along with the bound-exciton case.
The force differences agree between the two for atoms close to the defect.
Further away, the slow polynomial decay of the monopole moment from the change in charge state deviates from the exponential decay of the bound exciton.
(We note that increasing $\sigma_{BE}$ produces a larger region over which the force differences agree, as expected.)
We suggest applying a radial cutoff to the force difference obtained from a change in charge state as a practical approach to approximate the long-range behavior of the bound exciton.
This approach is further justified \textit{a posteriori} based on our analysis of luminescence in \cref{sec:lum}.

In our formulation, the force difference in the BvK supercell is already effectively zero-padded (\cref{fig:bvk}).
The radial cutoff zeros out more forces, removing regions where the bound-exciton and change-in-charge-state force differences disagree.
An additional role that the radial cutoff may serve is to impose a radial symmetry on the force differences and the associated long-range displacements.

\begin{figure}[htb!]
    \centering
    \includegraphics[width=\columnwidth,height=0.3\textheight,keepaspectratio]{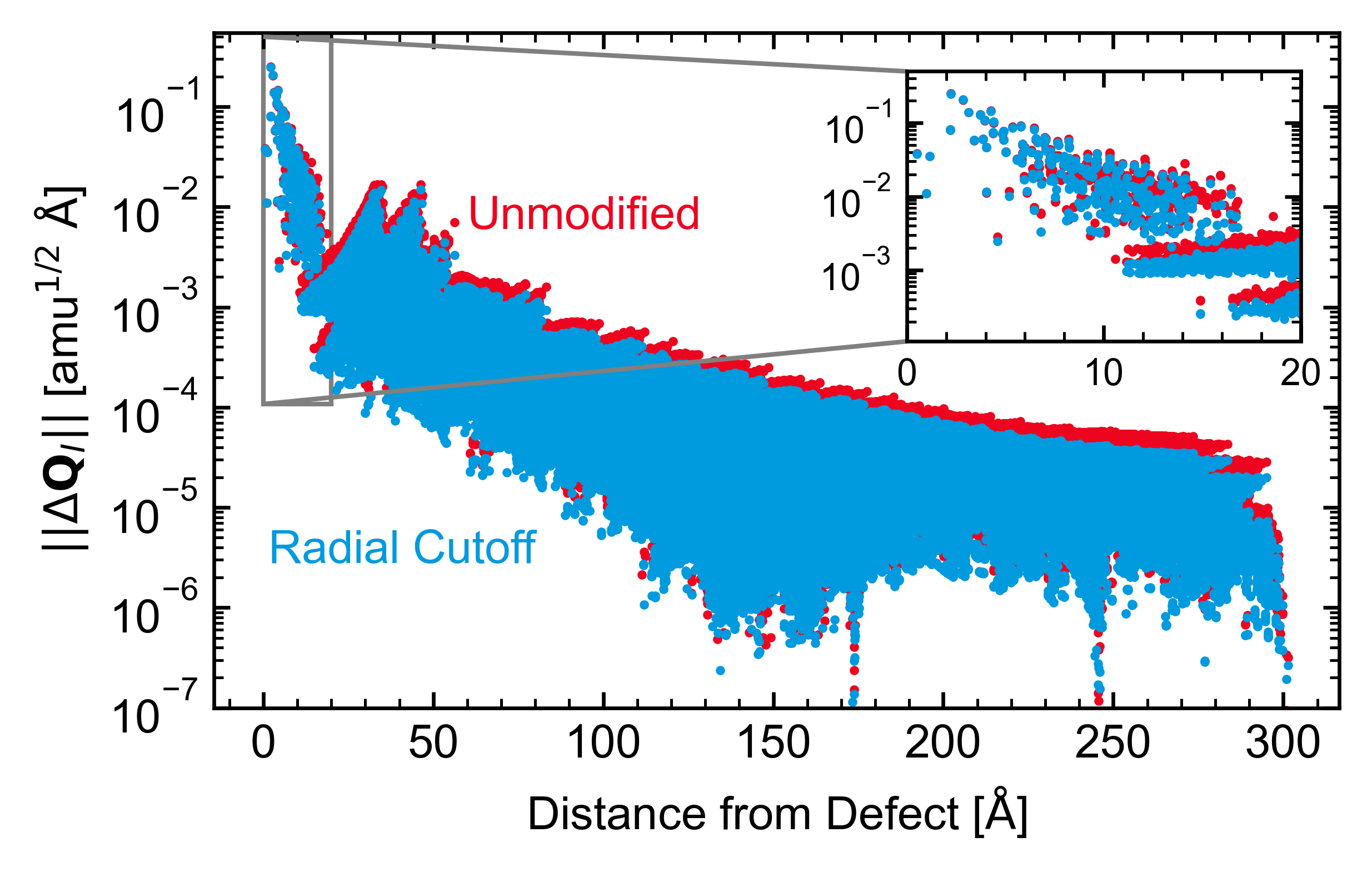}
    \caption{\label{fig:rad_cut}
        The displacement magnitude $\lVert \dQ_I \rVert$ on the $I$th atom as a function of distance from the defect in the {\bvk} supercell without (red) and with (blue) a radial cutoff.
        We take the average position of the C and H atoms in the T center as the center of the defect.
    }
\end{figure}

The atomic displacements $\dQ^{\rm BvK}$ resulting from the radial cutoff on the force differences are shown in \cref{fig:rad_cut}.
In general, the displacements are very similar.
Displacements close to the defect are identical to without the cutoff, while those further from the defect are slightly suppressed.
The cosine similarity between the two displacement vectors is 0.96, indicating high overlap.

\section{Luminescence Theory with Brillouin-Zone Sampling}
\label{sec:lum}
We evaluate the luminescence spectra of the NV center in diamond and the T center in Si using our formulation to obtain dilute-limit defect displacements.
To achieve this, we generalize the generating-function approach of Alkauskas \textit{et al.}~\cite{alkauskas_first-principles_2014-1,alkauskas_first-principles_2012}, which is based on the works of Lax~\cite{lax_franckcondon_1952} and Kubo and Toyozawa~\cite{kubo_method_1954,kubo_application_1955}, to include {\qpt} sampling.
Temperature dependence is also included~\cite{jin_photoluminescence_2021}.

The per-mode atomic displacements $\tjq$ are the essential quantities for evaluating luminescence.
From these, we define a partial Huang-Rhys factor given by
\begin{equation}
    \label{eq:Sj}
    S_{j {\bm q}} = \frac{1}{2\hbar} w_{\bm q} \, \omega_{j {\bm q}} {\lvert \tjq \rvert}^2 \;,
\end{equation}
where we've included the {\qpt} weight $w_{\bm q}$ in the definition.
By doing so, $S_{j {\bm q}}$ retains its interpretation as the average number of phonons with frequency $\omega_{j {\bm q}}$ emitted during the transition.

The total Huang-Rhys factor $S_{\rm tot}$ is used to quantify the strength of electron-phonon coupling.
It is the sum of partial Huang-Rhys factors,
\begin{equation}
    \label{eq:Stot}
    S_{\rm tot} = \sum_{j,{\bm q}} S_{j {\bm q}} \;.
\end{equation}
$S_{\rm tot}$ is used to define the Debye-Waller factor $e^{-S_{\rm tot}}$, which describes the fraction of light emitted into the zero-phonon line (as opposed to the phonon sideband) in luminescence.

\begin{figure}[htb!]
    \centering
    \includegraphics[width=\columnwidth,height=0.3\textheight,keepaspectratio]{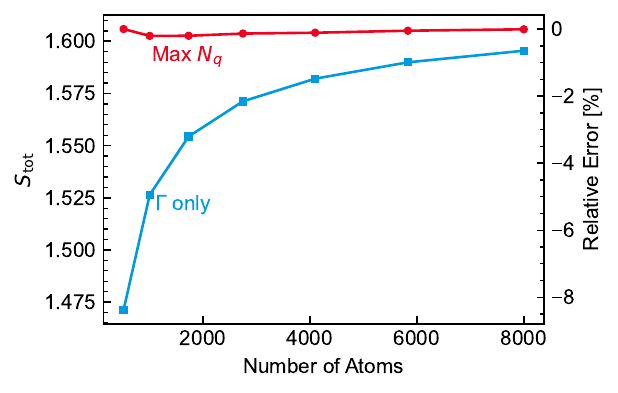}
    \caption{\label{fig:s_conv}
        Total Huang-Rhys factor $S_{\rm tot}$ [\cref{eq:Stot}] as a function of the number of atoms in the embedded supercell size.
        Results for $\Gamma$ point-only sampling are shown in blue and for the largest {\qpt} grid in red.
        Error is measured relative to the largest supercell with the max {\qpt} grid.
    }
\end{figure}

\Cref{fig:s_conv} shows the convergence of $S_{\rm tot}$ for the T center in Si as a function of the embedded supercell size (the equivalent of \cref{fig:dq_conv}).
Our method for {\qpt} sampling drastically accelerates convergence compared to $\Gamma$-only sampling.
Here we observe that $S_{\rm tot}$ (\cref{fig:s_conv}) converges more rapidly compared to $\Delta Q$ (\cref{fig:dq_conv}).
$\Delta Q$ is particularly sensitive to the low-frequency acoustic phonon modes, but these modes make a comparatively small contribution to $S_{\rm tot}$ because $S_{j {\bm q}} \propto \omega_{j {\bm q}}$.
$S_{\rm tot}$ from the smallest supercell is already within a fraction of a percent of the converged value.

We define the spectral density of electron-phonon coupling $S(\hbar\omega)$ as
\begin{equation}
    \label{eq:Shw}
    S(\hbar\omega) = \sum_{j,{\bm q}} S_{j {\bm q}} \, \delta (\hbar\omega - \hbar\omega_{j {\bm q}}) \;,
\end{equation}
and the temperature-dependent counterpart
\begin{equation}
    \label{eq:Chw}
    C(\hbar\omega, T) = \sum_{j,{\bm q}} \bar{n}_{j {\bm q}} (T) \, S_{j {\bm q}} \, \delta (\hbar\omega - \hbar\omega_{j {\bm q}}) \;,
\end{equation}
where $\bar{n}_{j {\bm q}} (T) = {\left[ \exp (\hbar\omega_{j {\bm q}} / k_{\rm B} T) - 1 \right]}^{-1}$ is the average occupation factor of the $j$th mode at temperature $T$.
The final steps to obtain the normalized luminescence intensity from our modified $S(\hbar\omega)$ and $C(\hbar\omega, T)$ proceed as usual.
We refer the reader to \cref{app:lum} for the remaining equations, as well as Refs.~\onlinecite{alkauskas_first-principles_2014-1,jin_photoluminescence_2021,turiansky_machine_2026}.

The delta functions in \cref{eq:Shw,eq:Chw} are often replaced with Gaussian distributions to produce a smooth spectral function.
A single broadening parameter may be introduced~\cite{alkauskas_first-principles_2014-1} or the broadening may vary linearly with the phonon frequency~\cite{razinkovas_vibrational_2021,jin_photoluminescence_2021}.
When local vibrational modes are present, a Lorentzian may be used in place of the Gaussian~\cite{silkinis_optical_2025,turiansky_machine_2026}.
Here we utilize Lorentzian broadening for \textit{all} modes.
The use of a Lorentzian is intended to suggest that genuine phonon lifetimes could be used, enabled by the high sampling density provided by our approach.
Evaluating the phonon lifetimes from first principles is computationally demanding---even with the acceleration of a MLIP---due to the combinatorics.
It is beyond the scope of our work to implement this idea, but is a promising direction for future exploration.
For the NV center, the half-width at half maximum is linearly varied from 2.0~meV at zero frequency to 0.5~meV at the maximum phonon frequency, and for the T center, it is varied from 0.25~meV to 0.1~meV.

\begin{figure}[htb!]
    \centering
    \includegraphics[width=\columnwidth,height=0.3\textheight,keepaspectratio]{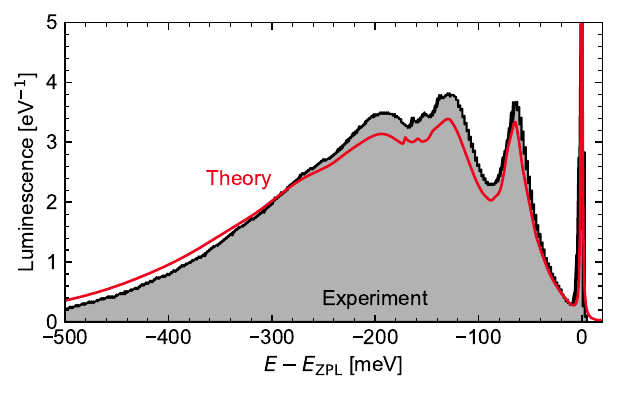}
    \caption{\label{fig:nv}
        Calculated luminescence spectrum of the NV center in diamond at $T = 10$~K (red).
        Experimental data (black) was extracted from Ref.~\onlinecite{kehayias_infrared_2013}.
    }
\end{figure}

The NV center is the prototype defect used for calculations of defect luminescence~\cite{alkauskas_first-principles_2014-1}.
Our calculated luminescence spectrum for the NV center is shown in \cref{fig:nv}.
Good agreement with the experimental spectrum of Ref.~\onlinecite{kehayias_infrared_2013} is obtained.
In particular, coupling to acoustic phonon modes near the zero-phonon line, which could be considered the hallmark of dilute-limit atomic displacements, is well captured.
Minor remaining differences could be attributed to, for example, the overestimation of $S_{\rm tot}$ compared to experiment (3.67 vs.\ 3.49), which is in part tied to the underlying exchange-correlation functional's description of the phonons~\cite{razinkovas_vibrational_2021}.
Our results for the NV center are also in good agreement with previous calculations using the embedding approach~\cite{alkauskas_first-principles_2014-1,razinkovas_vibrational_2021,jin_photoluminescence_2021}.
The main advantage of our method in the context of luminescence prediction is its efficiency.

\begin{figure}[htb!]
    \centering
    \includegraphics[width=\columnwidth,height=0.3\textheight,keepaspectratio]{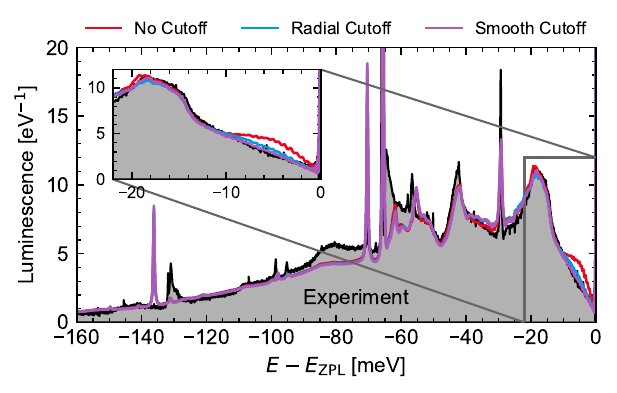}
    \caption{\label{fig:tcenter}
        Calculated luminescence spectrum of the T center in Si at $T = 4.2$~K without a radial cutoff (red), with a 9~{\AA} radial cutoff (blue), and with a smooth radial cutoff that varies linearly from 8~{\AA} to 9~{\AA} (purple).
        Experimental data (black) from Ref.~\onlinecite{tcenter} was measured as described in \cref{app:expt}.
    }
\end{figure}

\Cref{fig:tcenter} shows the calculated luminescence spectrum of the T center in Si.
Here we observe a distinct and unexpected enhanced coupling to acoustic phonons in the spectrum (see the inset of \cref{fig:tcenter}).
We attribute this enhancement to the incorrect long-range decay of the force difference $\dF$:
Recall that the excited-state geometry of the T center is approximated with the geometry of a charge state (\cref{app:comp_det}), so $\dF$ decays as a monopole (Case II) rather than exponentially, as expected for the true bound-exciton transition (Case III) discussed in \cref{sec:forces}.
Indeed, the internal transition of the NV center did not show this behavior, reinforcing the idea that this is unique to the bound-exciton physics.

As we suggested in \cref{sec:forces}, applying a radial force cutoff should instill the correct behavior.
We find a better description of the coupling to acoustic phonons with a radial cutoff of 9~{\AA} (see \cref{fig:tcenter}).
Our spectrum is in agreement with high resolution experimental measurements~\cite{tcenter} (details of the experiment can be found in \cref{app:expt}).
Empirically, we found that smoothly applying the radial cutoff, linearly decreasing the forces to zero between 8~{\AA} and 9~{\AA}, produced minorly improved agreement with experiment for coupling to low-frequency modes (\cref{fig:tcenter}).
The radial cutoff produces almost no change in the coupling to higher frequency modes.
A forthcoming work will investigate the minor remaining discrepancies with respect to experiment in detail.

\section{Dilute-Limit Configuration Coordinate Diagrams}
\label{sec:ccd}
Introduced in \cref{sec:bkgnd}, CCDs are an essential tool for analyzing transitions at defects.
Extrapolation of the transition energy $\Delta E$ to the dilute limit when there is a change in charge state is standard~\cite{freysoldt_fully_2009,freysoldt_first-principles_2014,komsa_finite-size_2012}.
With the invention of vertical charge-state corrections~\cite{gake_finite-size_2020,falletta_finite-size_2020}, it was suggested that the accepting-mode frequencies in CCDs should be corrected~\cite{kumagai_finite-size_2023}.
While the significant dependence of $\Delta Q$ on supercell size has been observed (e.g., in Ref.~\onlinecite{kumagai_finite-size_2023}), the dilute-limit scaling of $\Delta Q$ has been largely unaddressed.
Our work provides a clean and efficient way to extrapolate $\dQ$ to the dilute limit, and the values we obtain for the NV center and T center are given in \cref{tab:ccd}.
With $\dQ$ addressed, this approach provides the final ingredient necessary to obtain truly dilute-limit CCDs.

\begin{table}[ht!]
    \centering
    \caption{\label{tab:ccd}
        Magnitude of the atomic displacement $\Delta Q$ and accepting-mode frequency $\Omega_{\rm A}$ as explicitly calculated in a 512-atom supercell and in the dilute limit, as obtained by our procedure.
        The values for the T center include a smooth radial cutoff on the forces, as discussed in \cref{sec:lum}.
        For this reason, the T center's dilute-limit $\Delta Q$ value differs from \cref{fig:qpt_conv}.
    }
    \begin{ruledtabular}
        \begin{tabularx}{\columnwidth}{c c c c c}
            & \multicolumn{2}{c}{$\Delta Q$ [amu$^{1/2}$ {\AA}]} & \multicolumn{2}{c}{$\Omega_{\rm A}$ [meV]} \\
            Defect & 512 atom & BvK & 512 atom & BvK \\
            \midrule
            NV in diamond&0.694&0.768&65.4&59.3 \\
            T Center in Si&0.664&0.762&32.3&28.3 \\
        \end{tabularx}
    \end{ruledtabular}
\end{table}

What has not been appreciated in the literature is that not only does the magnitude of $\dQ$ change, but also its direction.
In addition to the long-range displacements that are present in the dilute limit, there are small adjustments to the displacements within the origin cell (\cref{fig:big_dq}).
Both of these effects mean that the effective frequency along $\dQ$ (the accepting-mode frequency) also changes in the dilute limit.
This physics is distinct from the change in accepting-mode frequency $\Omega_{g/e}$ suggested in Ref.~\onlinecite{kumagai_finite-size_2023}, which only arises in polar materials and is addressed in more detail in \cref{sec:discuss}.
Previously, we have provided a formula to obtain $\Omega_{g/e}$ directly from the dynamical matrix in lieu of a finite-difference evaluation [Eq.~(17) in Ref.~\onlinecite{turiansky_approximate_2025}].
Generalizing that formula to include {\qpt} sampling, we define the accepting-mode frequency $\Omega_{\rm A}$ as
\begin{equation}
    \label{eq:acc_mode}
    \Omega_{\rm A}^2 = \frac{1}{\Delta Q^2} \sum_{j,{\bm q}} w_{\bm q} \, \omega_{j {\bm q}}^2 \, {\lvert \tjq \rvert}^2 \;.
\end{equation}
Evaluating this expression with $\tjq$ from \cref{eq:tau_q} gives the dilute-limit accepting-mode frequency.
Whether $\Omega_{\rm A}$ corresponds to the frequency of the ground ($\Omega_g$) or excited ($\Omega_e$) state depends on which phonon modes (and therefore $\tjq$) are used.
(Recall that we are using the ground-state phonons here, which is discussed further in \cref{sec:discuss}.)

\begin{figure}[htb!]
    \centering
    \includegraphics[width=\columnwidth,height=\textheight,keepaspectratio]{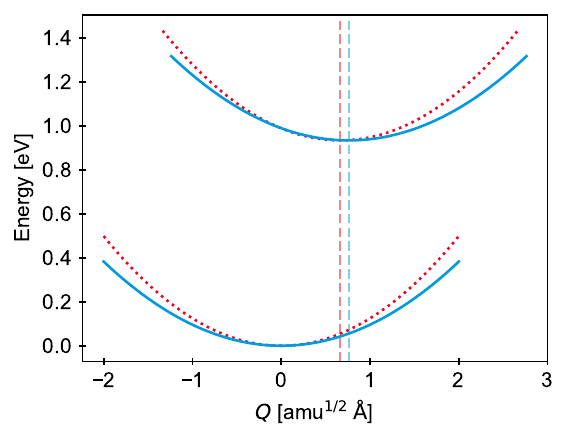}
    \caption{\label{fig:t_ccd}
        A configuration coordinate diagram for the T center in silicon calculated in a 512-atom cell (red, dotted) and in the dilute limit (blue, solid).
        Vertical dashed lines indicate the minimum of the excited-state potential energy surface.
        The equal-mode approximation is assumed here, and the zero-phonon line energy is the experimental value of 935~meV~\cite{bergeron_silicon-integrated_2020}.
    }
\end{figure}

Our results for the accepting-mode frequency are shown in \cref{tab:ccd}, and the CCD for the T center in silicon is shown in \cref{fig:t_ccd}.
In both cases, the dilute-limit frequency is reduced by approximately 10\% with respect to the 512-atom calculation.
Long-range displacements are primarily driven by coupling to acoustic phonon modes, which have low frequencies.
As the accepting-mode frequency is obtained by averaging all phonon modes weighted by their displacement, incorporating acoustic phonon modes drives down the average frequency.
A reduction in frequency of this magnitude can be quite significant.
For example, the nonradiative transition rate is exponentially dependent on the barrier energy determined by the crossing point between the two PESs~\cite{huang_theory_1950,englman_energy_1970,stoneham_theory_1975}.
Assessing the impact of our methodology for nonradiative decay is an important direction for future work.

\section{Discussion}
\label{sec:discuss}
\begin{figure}[htb!]
    \centering
    \includegraphics[width=\columnwidth,height=\textheight,keepaspectratio]{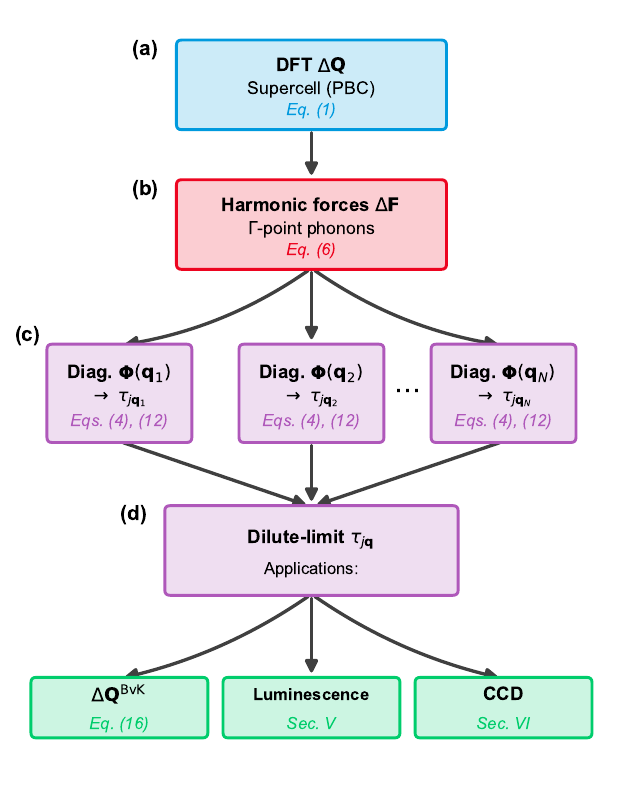}
    \caption{\label{fig:workflow}
        The workflow of our proposed procedure: (a) the calculated $\dQ$ [\cref{eq:dq}] is (b) transformed to a harmonic force difference $\dF$ [\cref{eq:dF}], followed by (c) parallel diagonalization of each ${\bm q}$-dependent dynamical matrix [\cref{eq:diag}] to obtain the per-mode displacements $\tjq$ [\cref{eq:tau_q}] for (d) dilute-limit applications.
    }
\end{figure}

The general workflow of our proposed procedure is depicted in \cref{fig:workflow}.
$\dQ$ is first calculated with DFT in a supercell within periodic boundary conditions [\cref{fig:workflow}(a)].
Using the $\Gamma$-point phonons, $\dQ$ is transformed into harmonic forces $\dF$ through \cref{eq:dF} [\cref{fig:workflow}(b)].
In parallel, each ${\bm q}$-dependent dynamical matrix in the {\qpt} grid is diagonalized [\cref{eq:diag}], and the per-mode displacements are obtained through \cref{eq:tau_q} [\cref{fig:workflow}(c)].
These per-mode displacements $\tjq$ have effectively transformed the explicitly calculated $\dQ$ into the dilute limit, where various quantities like the full $\dQ^{\rm BvK}$ [\cref{eq:big_dq}], luminescence (\cref{sec:lum}), or a CCD (\cref{sec:ccd}) can be obtained [\cref{fig:workflow}(d)].
We anticipate the utility of our formulation to extend beyond the applications demonstrated here.

We now discuss several important approximations used in and extensions of our approach: the influence of anharmonicity, the equal-mode approximation, the cost of evaluating the dynamical matrix, and the treatment of polar materials.

\textit{Anharmonicity--- }
Throughout this work, we have relied on a harmonic approximation.
Anharmonicity is often present in the PES of defects and can be quite important for nonradiative carrier capture~\cite{stoneham_non-radiative_1981,alkauskas_first-principles_2014,kim_anharmonic_2019,zhang_iodine_2020,zhang_minimizing_2021,kavanagh_rapid_2021,zhang_defect_2022,dou_chemical_2023}.
The harmonic approximation is used in the conversion from DFT displacements $\dQ$ into forces $\dF$; essentially, we are asking what forces would produce this displacement pattern for a harmonic PES.
Those forces are then transformed back into displacements in the BvK cell, and we found that the displacements close to the defect, which are the largest in magnitude, are maintained (\cref{fig:big_dq}).
The largest displacements are those for which potential differences due to anharmonicity may be expected to arise, and those displacements match the true DFT calculation, which includes anharmonicity.
The long-range displacements are quite small and are therefore expected to be well described by a harmonic assumption.
Based on these two facts, our procedure is applicable to cases with anharmonicity.

However, one consequence of anharmonicity is that $\dF$ from \cref{eq:dF} may no longer be exactly the same as the $\dF$ evaluated explicitly with DFT via \cref{eq:dF_def}.
Our analysis of the long-range decay of $\dF$ by explicit calculation in \cref{sec:forces} loses an exact mapping onto the harmonic $\dF$ used in our procedure.
Even so, this analysis was largely based on model densities, which are primarily determined by the localization of the states involved in the transition.
For an anharmonic PES, the essential question becomes whether the states involved maintain their qualitative localization over the relevant range of displacements from the equilibrium structure.
If so, one can reasonably expect the long-range decay of the force differences to be maintained whether or not anharmonicity is present, thus validating our analysis.

\textit{Equal-Mode Approximation--- }
We now relax the equal-mode approximation that we have relied on.
Under a different dynamical matrix (${\bm \Phi}_e$ instead of ${\bm \Phi}_g$) but the same $\dQ$ [\cref{eq:dq}], \cref{eq:dF} will produce a \textit{different} $\dF$.
The phonon eigenvectors $\etajq$ from the new dynamical matrix will then be used to convert back into per-mode displacements $\tjq$ in the BvK supercell [\cref{eq:tau_q}].
As we found in \cref{sec:unfold}, displacements near the defect are exactly reproduced under this procedure by construction.
Long-range displacements are a harmonic response of the bulk lattice and can be reasonably assumed to stay the same under the excited-state dynamical matrix:
differences in the excited state are likely due to the local electronic structure of the defect and not some difference of the bulk electronic structure.
Thus, while strictly speaking the dilute-limit displacement $\dQ^{\rm BvK}$ may differ, we can reasonably expect this procedure to produce a very similar $\dQ^{\rm BvK}$ under the excited-state dynamical matrix.

While $\dQ^{\rm BvK}$ can be expected to be the same, the per-mode displacements $\tjq$ will differ.
Differences in $\tjq$ reflect the differing phonon eigenvectors $\etajq$ in the ground and excited state.
For processes that rely directly on $\tjq$, the most relevant dynamical matrix should be chosen.
In an optical transition, this corresponds to the final state: the ground state should be chosen for luminescence, as we did here, and the excited state for absorption.
In the context of CCDs, the accepting-mode frequency $\Omega_{\rm A}$ calculated through \cref{eq:acc_mode} will differ in the ground and excited state.
This difference reflects genuine physics from the differing PESs.

\textit{Cost of the Dynamical Matrix--- }
The need to obtain the full dynamical matrix of the defective supercell may seem like a limitation of our approach---particularly if it is needed for both the ground and excited state.
This potential limitation is most obvious for CCDs, where a simple CCD can be constructed with just two fixed-geometry calculations displaced from the equilibrium geometries of each state.
In previous work~\cite{turiansky_machine_2026}, we demonstrated that the dynamical matrix can be obtained with hybrid-functional DFT accuracy by fine-tuning a MLIP to the atomic relaxation data generated in the determination of the equilibrium geometry.
The training time takes less than the fixed-geometry calculations and provides access to the full dynamical matrix, not only the accepting-mode frequency for the CCD.
Other approaches are possible~\cite{mannodi-kanakkithodi_accelerating_2026,zheng_compressive-sensing-enhanced_2024}.
In our opinion, obtaining the dynamical matrix is a rather negligible cost with modern techniques.

\textit{Polar Materials--- }
The materials explored in this work, Si and diamond, do not contain polar phonon modes.
Previous works have relied on an assumption of short-ranged interactions to construct dynamical matrices of the large, embedded supercells~\cite{alkauskas_first-principles_2014-1,razinkovas_vibrational_2021}, which limits the applicability to materials without polar phonon modes.
This is, in our opinion, a major advantage of our approach.
Non-analytical corrections to the ${\bm q}$-dependent dynamical matrix to account for long-range interactions are standard~\cite{pick_microscopic_1970,gonze_dynamical_1997,togo_implementation_2023}.
While this is a straightforward extension to our work that enables addressing materials with polar phonon modes, we did not pursue it here as it requires evaluating Born effective charges of the defective supercell, which is expensive.
Future work will explore this route, in particular to identify strategies to minimize the cost of Born effective charge evaluation.

Our analysis in \cref{sec:forces} shows that the long-range behavior of the force difference $\dF$ is independent of the presence of polar modes.
This result was surprising to at least one of the Authors but follows plainly from the force definition in DFT.
Ultimately it is the electron density difference $\delta n$ of the transition that dictates the long-range behavior.
Even if both states of the transition are charged---as exemplified by the NV center---and there is contamination of the respective equilibrium geometries ${\bm R}^0_{g/e}$ due to interactions with periodic images, the associated $\dF$ is unaffected.
Instead, that contamination may impact the phonon modes of the system.
As shown in \cref{fig:qpt_conv} and through the good agreement with respect to experiments in \cref{sec:lum}, this contamination is a relatively minor effect.

\textit{Vertical Charge-State Correction--- }
In a vertical transition between two states of a defect, the atomic coordinates are kept fixed.
As only electronic degrees of freedom are allowed to relax, the long-range interactions with periodic images of the defect only experience screening by the high-frequency dielectric permittivity~\cite{falletta_finite-size_2020}.
This is in contrast to the transition energy $\Delta E$, which is evaluated with the total energies of each state in its respective equilibrium geometry where both ionic and electronic screening can occur, requiring the use of the static dielectric permittivity to evaluate screening.
On the basis of the difference in screening as the atoms are displaced, a correction to the accepting-mode frequency has been suggested~\cite{kumagai_finite-size_2023}, which we will refer to as the polar accepting-mode correction.
This correction tends to increase the accepting-mode frequency.

However, the static and high-frequency dielectric permittivities of nonpolar materials, like Si and diamond, are identical, so no correction is needed.
Even so, we observed a significant suppression of the accepting-mode frequency, which we attributed to coupling to acoustic modes in the dilute limit.
One concludes that there may be a fortuitous cancellation of errors in the accepting-mode frequencies that are evaluated in small supercells of polar materials.
This may be why calculations of nonradiative rates and other properties that rely on CCDs have been successful in describing experiments~\cite{alkauskas_first-principles_2014,turiansky_nonrad_2021,zhang_minimizing_2021,wilhelmer_optical_2025,kavanagh_rapid_2021}, despite neglecting to address these two effects.
On the other hand, $\Delta Q$ still can depend significantly on supercell size, even if there is a cancellation in the accepting-mode frequency.

While the polar accepting-mode correction is useful for CCDs, and therefore nonradiative transitions, it leaves questions regarding the other $3N-1$ degrees of freedom in the system.
As our luminescence examples show, this is particularly relevant.
The accepting mode is not itself a phonon mode of the system, having contributions from various phonon modes.
Non-analytical corrections to the ${\bm q}$-dependent dynamical matrix accelerate the convergence of the phonon modes of polar materials.
It is not clear whether these non-analytical corrections address the same physics as the polar accepting-mode correction.
If not, perhaps the vertical-transition correction could be reformulated in ${\bm q}$-space, providing a path to addressing all phonon modes of the system.
We found through the dual embedding/{\qpt}-sampling approach (\cref{fig:dq_conv,fig:s_conv}) that the phonon modes converge rapidly with supercell size, indicating that the perturbation from the defect itself is minor, at least in nonpolar materials.

\section{Conclusion}
\label{sec:concl}
We have presented an approach for describing the transition-induced atomic displacements of defects in the dilute limit by exploiting Brillouin-zone sampling.
In this approach, atomic displacements are obtained in small supercells within periodic boundary conditions and converted into a force difference under a harmonic PES, which is then unfolded into the {\bvk} supercell.
This procedure yields per-mode atomic displacements that describe the minute displacements on atoms hundreds of {\AA}ngstr{\"o}ms from the defect, which are inaccessible by direct calculation.
Compared to the established embedding approach, our method converges to the dilute limit far more rapidly in terms of supercell size, requires the diagonalization of only small dynamical matrices, and efficiently parallelizes over {\qpt}s.

We carefully analyzed the localization behavior of the force difference, which we show is generated from the electron density difference between the ground and excited state.
By considering different possible model densities in both periodic and open boundary conditions, we demonstrated that these force differences are sufficiently short-ranged for internal transitions and---provided extra care is exercised---for bound-exciton transitions.
These results justify the use of forces computed in modest supercells, both within our formalism and within embedding.

Applying our formalism, we obtained luminescence spectra of the NV center in diamond and T center in Si that compared favorably with experiments.
In particular, we were able to capture coupling to acoustic phonon modes, which is the hallmark of dilute-limit displacements.
In addition, we applied our approach to obtain dilute-limit CCDs.
Not only do we obtain the magnitude of the atomic displacement but also the accepting-mode frequency, which decreases in magnitude due to acoustic-phonon coupling.

These results provide the missing ingredients necessary to truly address defect transitions in the dilute limit.
We anticipate broad utility beyond the applications demonstrated here, for example in nonradiative capture rates and other properties that hinge on the change in geometry between defect states.
Promising extensions include incorporating non-analytical corrections to treat polar materials and using genuine phonon lifetimes in the luminescence calculation, enabled by the high sampling density of our approach.

\begin{acknowledgments}
    We gratefully acknowledge fruitful discussions with M.\ Kazemi, M.\ Keshavarz, D.\ B.\ Higginbottom, and M.\ L.\ W.\ Thewalt.
    This work was supported by the Office of Naval Research through the Naval Research Laboratory's (NRL) Basic Research Program and in part by an appointment to the U.S. Naval Research Laboratory Research Transitional Postdoctoral Program, administered by the Office of Fellowships of the National Academies of Sciences, Engineering, and Medicine.
    Calculations were supported by high-performance computer time and resources from the DoD High Performance Computing Modernization Program.
\end{acknowledgments}

\section*{Data Availability}
The data that supports the findings of this study are available via Github at \url{https://github.com/mturiansky/lineshape_tools} (Ref.~\onlinecite{turiansky_lineshape_tools_2025}) in the \verb|examples/| directory.

\section*{Code Availability}
The \verb|lineshape_tools| code (v0.2) implements this methodology and is available via Github at \url{https://github.com/mturiansky/lineshape_tools} (Ref.~\onlinecite{turiansky_lineshape_tools_2025}).

\appendix

\section{Computational Details}
\label{app:comp_det}
DFT calculations are performed with the VASP code~\cite{kresse_efficient_1996,kresse_efficiency_1996} version 6.4.3.
We freeze the core electrons within the projector augmented-wave formalism~\cite{blochl_projector_1994}.
The valence electrons are represented in a plane-wave basis truncated at an energy of 400~eV for Si and 520~eV for diamond.
We utilize the HSE06 hybrid functional~\cite{heyd_hybrid_2003,*heyd_erratum:_2006} (with the default mixing parameter of 0.25 and screening parameter of 0.2~{\AA}$^{-1}$) for diamond, consistent with previous studies~\cite{alkauskas_first-principles_2014-1,gali_ab_2019}.
For Si, we utilize the PBE0 hybrid functional~\cite{perdew_rationale_1996,adamo_toward_1999} with a mixing parameter of 13.6\%, as it has been found empirically to predict transition energies of the T center closer to experiments when extrapolated to the dilute limit~\cite{nangoi_carbon-nitrogen_2026}.

Following the usual approach~\cite{freysoldt_first-principles_2014}, defects are simulated in a supercell within periodic boundary conditions.
For both systems, we construct a supercell from a $4\times4\times4$ multiple (512-atom) of the conventional cubic unit cell.
Details of the atomic and electronic structure of the NV center~\cite{gali_ab_2019} and T center~\cite{bergeron_silicon-integrated_2020,dhaliah_first-principles_2022} are given elsewhere.

For electronic structure calculations of the supercells, the Brillouin zone is sampled at the $\Gamma$ point.
Atomic coordinates are relaxed until the forces are below 1~meV/{\AA}, while the lattice parameters are kept fixed at the calculated bulk values.
The excited state of the NV center in diamond is simulated using constrained-occupation $\Delta$SCF~\cite{jones_density_1989}.
We use the low-symmetry occupations of the NV center excited state, resulting in a Jahn-Teller distortion.
A complete analysis of NV center luminescence should account for the Jahn-Teller effect~\cite{bersuker_jahn-teller_2006,razinkovas_vibrational_2021,zalandauskas_theory_2025}, but this is beyond the scope of the present study.
The excited state of the T center involves a bound exciton whose geometry is approximated with that of the negative charge state, following previous work~\cite{silkinis_optical_2025,turiansky_machine_2026}.
We discuss some implications of this choice for the T center in more detail in \cref{sec:forces}.

To avoid the significant expense of evaluating the dynamical matrix (see \cref{sec:lum}) of large supercells, we utilize the acceleration provided by machine learning interatomic potentials (MLIPs) with the procedure described in Ref.~\onlinecite{turiansky_machine_2026}, as implemented in the \verb|lineshape_tools| code (v0.2)~\cite{turiansky_lineshape_tools_2025}.
For our MLIP framework, we employ the 64-bit floating point implementation of the atomic cluster expansion with message passing (MACE)~\cite{batatia_mace_2022,batatia_design_2025}.
We utilize the \verb|mace-omat-0-medium| foundation model~\cite{batatia_foundation_2025}, which is pre-trained on the Open Materials 2024 dataset~\cite{barroso-luque_open_2024}, as the starting point for fine-tuning.
Following the suggestion of Ref.~\onlinecite{turiansky_machine_2026}, we fine-tune~\cite{tompa_fine-tuning_2026} to a dataset containing (i) atomic relaxation data generated to obtain the equilibrium geometry of the given defect, (ii) 10 randomly perturbed configurations of the pristine, bulk supercell, and (iii) a few additional configurations.
Our relaxation datasets contain 28 configurations with a maximum force on any atom of 9.8~eV/{\AA} for the NV center and 12 configurations with a maximum force of 1.0~eV/{\AA} for the T center.
We use the optimized phonon method~\cite{turiansky_machine_2026} to generate the 30 additional configurations for each defect.
The T center has prominent local vibrational modes in its spectrum~\cite{safonov_interstitial-carbon_1996,bergeron_silicon-integrated_2020,turiansky_machine_2026}:
we explicitly include a configuration displaced along each local vibrational mode predicted by the foundation model, to ensure their energy is properly described.
Here local vibrational modes are identified as having an inverse participation ratio less than 10, of which there are 8.

\section{Alternative Choices for Incorporating Brillouin Zone Sampling}
\label{app:alt}
There are several different forms of the per-mode atomic displacements $\tjq$ that could be considered when performing {\qpt} sampling.
The simplest approach is to treat $\dQ$ as a $\Gamma$-point phonon mode (without converting into forces).
When unfolding into the BvK supercell, $\dQ$ is repeated in each image [\cref{fig:bvk}(a)] and multiplied by the phase factor [\cref{fig:bvk}(b)], which is just unity for the $\Gamma$ point.
However, since the phonon eigenvectors $\etajq$ are also multiplied by the phase factor in the BvK supercell (forming $\nujq$), $\tjq$ will be non-zero only for the $\Gamma$ point when integrating over the BvK supercell.
As a result, there is nothing gained by performing {\qpt} sampling with this approach.
One comes to a similar conclusion if the force difference $\dF$ is used and treated with a $\Gamma$-point phase factor.

One of these two choices is likely the origin of the following statement in Ref.~\onlinecite{razinkovas_vibrational_2021}:
``Because the relaxations have the periodicity of the lattice the problem is not resolved by calculating the phonon spectrum in the entire Brillouin zone of the supercell, as only $\Gamma$ phonons contribute.''
We agree that neither of these choices are useful, but as our work shows, there are other choices to be made.

Another choice, which is more fruitful but not optimal, is to treat $\dQ$ as a local vibrational mode.
Local vibrational modes have the same phonon eigenvector $\etajq$ for all {\qpt}s in the Brillouin zone up to an arbitrary phase factor.
Unfolding into the BvK supercell can proceed as usual [\cref{fig:bvk}(b)], but the phase factor of $\dQ$ can always be chosen to match that of $\etajq$.
As a result, $\tjq$ can be evaluated in the defect supercell as
\begin{equation}
    \label{eq:tau_q_bad0}
    \tjq^\textrm{non-optimal} = \sum_{I,\alpha} \eta_{j{\bm q};I\alpha}^* \, {\Delta Q}_{I\alpha}
\end{equation}
or in matrix form, ${\bm \tau}^\textrm{non-optimal} ({\bm q}) = {\bm U}^\dagger ({\bm q}) \, \dQ$.
Because $\dQ$ is used directly, the long-ranged displacements beyond the defect supercell are not captured.
For that reason, it is not optimal; however, this approach has the advantage of providing a better description of the phonon density of states compared to $\Gamma$-only sampling.
We are aware of at least one work~\cite{ivanov_effect_2022} that utilized this approach.

Instead of operating on $\dQ$ directly, one could use the forces:
\begin{equation}
    \label{eq:tau_q_bad1}
    \tjq^\textrm{non-optimal} = \frac{1}{\omega_{j {\bm q}}^{2}} \sum_{I,\alpha} \eta_{j {\bm q}; I \alpha}^* \, \Delta F_{I \alpha} \;,
\end{equation}
or in matrix form, ${\bm \tau}^\textrm{non-optimal} ({\bm q}) = {\bm \Omega}^{-2} ({\bm q}) \, {\bm U}^\dagger ({\bm q}) \, \dF$.
This choice also improves the description of the phonon density of states.
It fails to capture coupling to acoustic phonons because the force difference is repeated in each image within the BvK supercell, and this perturbs the displacements, preventing the long-range displacements from building.

The final possible choice is the equivalent of \cref{eq:tau_q} but operating directly on $\dQ$ without converting to a force difference.
When unfolding into the BvK supercell, this amounts to retaining $\dQ$ inside one image and setting it to zero outside [\cref{fig:bvk}(c)].
This choice is clearly non-optimal because the long-range displacements are enforced to be zero.

\section{Obtaining the Luminescence Intensity}
\label{app:lum}
Following Refs.~\onlinecite{alkauskas_first-principles_2014-1,jin_photoluminescence_2021,turiansky_machine_2026} to obtain the normalized luminescence intensity, we define the generating function $G(t, T)$, where
\begin{multline}
    \label{eq:Gt}
    G(t, T) = \exp [ -i E_{\rm ZPL} t / \hbar + S(t) - S(0) \\
    + C(t, T) + C(-t, T) - 2 C(0, T) ] \;.
\end{multline}
$E_{\rm ZPL}$ is the energy of the zero-phonon line, which is equivalent to $\Delta E$ under the equal-mode approximation.
$S(t) = \int e^{i \omega t} S(\hbar\omega) \, d(\hbar\omega)$ is the Fourier transform of $S(\hbar\omega)$ [\cref{eq:Shw}], and similarly for $C(t, T)$ [\cref{eq:Chw}].
Next we introduce an intermediate spectral function $A(\hbar\omega, T)$, given by
\begin{equation}
    \label{eq:Ahw}
    A(\hbar\omega, T) = \frac{1}{2\pi} \int_{-\infty}^{\infty} e^{i\omega t - \gamma \lvert t \rvert} \, G(t, T) \, dt \;,
\end{equation}
where $\gamma$ is the homogeneous broadening, taking a value of 0.4~meV for the NV center and 1.18~$\mu$eV for the T center.
Finally, the normalized luminescence intensity is given by
\begin{equation}
    \label{eq:Lhw}
    L(\hbar\omega, T) = \mathcal{N}_L \, \omega^3 A(\hbar\omega, T) \;,
\end{equation}
where $\mathcal{N}_L^{-1} = \int \omega^3 A(\hbar\omega, T) \, d(\hbar\omega)$ is a normalization constant.

\section{T Center Luminescence Measurements}
\label{app:expt}
The photoluminescence spectrum of an ensemble of T centres was measured in Ref.~\onlinecite{tcenter} using an isotopically purified $^{28}$Si bulk sample, previously used in studies of the T centre~\cite{kazemi_giant_2026,bergeron_silicon-integrated_2020}.
The sample was irradiated with \SI{10}{\mega\electronvolt} electrons to a total dose of \SI{320}{\kilo\gray}, followed by annealing in boiling water for 24 hours.
It was subsequently annealed in hydrogen gas at \SI{300}{\celsius} for 30 minutes, \SI{350}{\celsius} for 30 minutes, \SI{400}{\celsius} for 60 minutes, and finally at \SI{440}{\celsius} for 30 minutes.
For measurement of the photoluminescence, the sample was mounted in a reflective pocket holder and immersed in liquid helium at a temperature of \SI{4.2}{\kelvin}.
It was excited using above-bandgap light from a \SI{980}{\nano\meter} diode laser, and the emission spectrum was recorded with a spectral resolution of \SI{12.4}{\micro\electronvolt} using a Bruker IFS 125 HR Fourier transform infrared spectrometer equipped with a CaF$_2$ beamsplitter and a liquid-nitrogen-cooled Ge diode detector.

\end{document}